\documentclass{elsart}
\usepackage{graphicx}

\def\bibcode#1{(\texttt{#1})}

\begin{document}
\begin{frontmatter}
\title{Light pollution at the Roque de los Muchachos Observatory}

\author{Pedani, M.\thanksref{email}}

\address{INAF-Centro Galileo Galilei, Po Box 565, S/C La Palma - 38700 
TF, Spain}
\thanks[email]{E-mail: pedani@tng.iac.es}

\begin{abstract}
Sky spectra were obtained from archival science frames taken with 
DoLoRes at the 3.58m Telescopio Nazionale 
Galileo with a wavelength range $\sim 3800 - 8000 \AA$ and 
resolution of $2.8 \AA$/pix and $3.6 \AA$/pix. 
Our spectra include all the important Sodium and Mercury light pollution 
lines 
and span a wide interval of azimuth and observing conditions, essential to 
disentangle environmental and seasonal effects. 
New sodium and mercury lines were also detected for the first time at the 
observatory. 
Light pollution from NaD$_{5892-8}$ emitted by the LPS lamps 
increased by a factor of $1.5 - 2$ with respect to the average values 
of 1998. At the same time, light pollution from Hg lines 
decreased by $\sim 40\%$ and reaches the 1998 levels 
only when observing toward the towns. 
The contribution of NaD$_{5892-8}$ from LPS lamps to sky background is $0.05 
- 0.10$mag at V-band and $0.07 -0.12$mag at R-band. 
Synthetic sky brightness measures calculated from our spectra at V, B 
and R bands are in good agreement with those 
of Benn \& Ellison (1998) if we take into account that our observations were 
done during 2003, seven years after the last sunspot minimum. 
The effects of the application of the Canary Sky Law are directly visible 
in the spectra as a $50\%$ dimming of the Hg 
light-polluting lines in the spectra taken after local midnight.
\end{abstract}

\end{frontmatter}

\section{Introduction}
The Observatorio del Roque de los Muchachos (ORM), located at 
La Palma in the Canary Islands is actually the largest 
European Observatory in the northern hemisphere. The site benefits from 
good sky 
transparency, and a high fractions of clear ($\sim 70\%$) and 
photometric nights ($\sim 60\%$) and a mean seeing of $0.76"$ 
(Munoz-Tunon et al. 1997). 
The ORM is located at $\sim 2300m$ altitude, close to the summit of a $2426m$ 
volcanic peak at longitude 
$17.9^\circ$W and latitude $+28.7^\circ$ and very close to the rim of a 
caldera. An inversion layer in the $1300-1700m$ height range, guarantees 
(though with many exception in winter) stable observing conditions 
during $3/4$ of the year. 
The relative proximity ($\sim 200$Km Eastward) of the Moroccan coast makes 
it possible 
that, expecially during the summer, dust from the Sahara desert blows over the 
island, increasing atmospheric extinction (typically $\sim 0.09mag$ at 
r$^\prime$-band).
To our knowledge, after Benn \& Ellison (1998, hereinafter BE98), who 
presented a low-resolution night-sky spectrum taken at WHT in 1991 
(see their Fig.1), no other 
works have been published on light pollution at ORM. 
Our spectra have about three times greater resolution than that of 
BE98; they span a wide range of environmental parameters and observing 
conditions and show all the important light pollution lines. 
During recent years the island of La Palma underwent a strong 
development of turistic resources with the construction of new 
hotels, roads and urban areas. Though a special Sky Law exists which 
establishes the general rules for public and private illumination, this 
growth inevitably 
led to an increase of the outdoor lighting with negative consequences for 
the light pollution at the observatory.
The aim of this paper is to give a comprehensive and up-dated view of the 
light pollution at ORM during $2003$.

\noindent
The organization of the paper is as 
follows: the sources of light pollution in La Palma are described in 
Sect.2; the observational data are summarized in Sect.3 and the analysis of 
the night-sky spectra is described in Sect.4. 
Spectrophotometric data are described in Section 5 and conclusions are 
summarized in Sect.6.

\section{Light pollution at ORM}
About $85,000$ people live in La Palma, mainly concentrated in $8$ 
small towns within $15$Km of the ORM. 
Given the altitude of the ORM, the line-of-sight over the sea has a 
radius of $\sim 180$Km, enough to intercept the lighting of the major 
Canary island Tenerife ($800000$ people and $120$Km distant) 
whose coast is visible to the naked eye on very clear nights.
Nevertheless, its contribution to the sky brightness, 
as well as that of two small islands, (El Hierro and La Gomera, $29000$ 
people and $40$Km distant) is negligible. 
In many cases, the presence of the so called "sea of clouds" below the 
thermal inversion layer, 
greatly reduces outdoor lighting, especially during the coldest months.
During summer, the presence of the anticyclone of Azores causes the clouds to 
be dispersed, so that outdoor lighting can easily escape upward.
The most important sources of light pollution in La Palma are listed in 
Table 1. 
Though the study of the sky brightness is not the aim of this paper, 
their contribution to the zenith sky brightness at V-band has been 
calculated using the model of Garstang (1989). This model (tested with 
some U.S. cities) is based on a series of assumptions which do not 
translate entirely to La Palma. 

\noindent
Though the ground reflectivity and 
the fraction of aerosols in the atmosphere can be those of a 
typical high-altitude site in the U.S. (e.g. Mount Graham), 
the fraction of outdoor lighting escaping upward is much less in La 
Palma. The relative fraction of lamps installed on La Palma 
(LPS lamps are much preferred here) is also different from other cities, 
so that light pollution preferentially arises toward red wavelenghts, 
with different impact on the sky brightness with respect to a 
site where mercury lamps are predominant. 
On the other hand, the above model assumes $1000$ Lumens/head 
which approximately agrees with the typical values of La Palma 
($\sim 1850$ Lumens/head before local midnight and $\sim 1000$ Lumens/head 
after). 

\noindent
The Canary Sky Law, introduced in $1992$ (McNally $1994$) put 
strict limits on the type of lamps which can be used for outdoor lighting, 
on their power, and on orientation with respect to 
the ground and implied that, after local midnight, most of the high-pressure 
sodium (HPS) and mercury lamps must be estinguished, as well as all 
the discharge-tube illumination. 
In general, LPS lamps should be used except in the urban areas where 
HPS lamps are admitted and a non-negligible fraction of mercury 
and incandescent lamps still exist (see Table 2).

\noindent
LPS lamps are the best choice for astronomy because 
their emission is almost exclusively concentrated in the NaD$_{5890-6}$ doublet, 
which simply adds to the natural sky glow at these wavelength. No 
continuum emission arises from these lamps. Other emission lines are 
Na$_{5683-8}$ and Na$_{6154-61}$, the latter about $4$ times weaker than 
the former. Detecting the above lines in the sky spectra permits the 
contributions to the NaD$_{5890-6}$ emission from light pollution and the 
natural sky 
glow to be disentangled (see Sect. 4.1).
Up to now, the only way to measure the natural NaD skyglow at ORM was during an 
artificial $1$hr blackout on the night $24-25$ June $1995$ to celebrate 
the $10th$ anniversary of the inauguration of the ORM (see BE98 for details). 

\noindent
The HPS lamps are the second contributor in terms of light output on La Palma 
(see 
Table 2). Their emission is characterized by a smooth continuum in the 
$\sim 5500$ to $7000 \AA$ range. The NaD$_{5890-6}$ line, is now replaced by a 
deep void. Other narrow emission line are: Na$_{4665-9}$, Na$_{4979-83}$, 
Na$_{5149-53}$, Na$_{5683-8}$ and Na$_{6154-61}$. 

\noindent
Mercury lamps, though they contribute with a mere $9\%$ to the total luminous 
flux of the island are another important source of light-polluting lines, 
expecially in the violet/blue region of the spectrum. There is also a weak 
continuum emission in the $3200 - 7800 \AA$ range. 
The most important lines observed in our spectra are: Hg$_{4046}$, 
Hg$_{4358}$, Hg$_{5461}$, Hg$_{5769}$ and Hg$_{5790}$ (see Sect. 4.2). 

\noindent
Incandescent lamps are a significant source of light pollution before 
midnight (see Table 2), though their solely continuum emission is not 
considered in the present work.
Nevertheless, BE98 estimated their contribution to zenith sky brightness 
at V-band to be $0.01$ mag.

\noindent
At La Palma, light pollution originates from $17166$ street lamps (end of year 
$2000$, $23\%$ more than reported in BE98) emitting 
a total of $1.56 \times 10^{5}~$KLumens before midnight, reduced to $1.0 \times 
10^{5}~$KLumens after that hour. If we consider that about $50\%$ of the light 
is emitted by the fixtures and the ground reflectivity is assumed $10\%$, 
we calculate that the amount of power emitted upward by the 
outdoor lighting is $\sim 16$W/Km$^{2}$ before midnight and $\sim 
11$W/Km$^{2}$ 
after. It is noteworthy that the typical sky background of $V = 21.9 
mag/arcsec^{2}$ corresponds to $\sim 9.2$W/Km$^{2}$.

%
\begin{table*}
\begin{center}
\caption[]{Sources of light pollution at the ORM at different azimuths (North 
through East). The 
contributions are calculated according to the model of Garstang (1989) and 
should be considered as upper limits (see Sect.2).}
\begin{tabular}{ccccc}
               &          &            &          &                   \\
\hline
Town           & Azimuth  & Population & Distance & $\Delta mag_{V}$ \\
               & $(deg.)$ & $(2003)$   & $(Km)$   &            \\
\hline
Barlovento     & $50$     & $2400$     & $10$     & $0.03$       \\
San Andres y Sauces & $110$    & $5100$     & $12$     & $0.05$    \\
Santa Cruz     & $125$    & $18200$    & $15$     & $0.13$       \\
Brena Alta/Baja & $140$   & $10800$    & $15$     & $0.07$       \\
El Paso        & $180$    & $7500$     & $12$     & $0.08$       \\
Los Llanos $+$ Tazacorte & $200$    & $26100$    & $12$     & $0.32$    \\
Puntagorda     & $280$    & $1800$     & $9$      & $0.03$       \\
Garafia        & $325$    & $2000$     & $9$      & $0.03$       \cr
\hline
\end{tabular}
\end{center}
\end{table*}
%

\begin{table*}
\begin{center}
\caption[]{Type and number of lamps installed at La Palma at the end of year $2000$ 
(Francisco Javier Diaz Castro - private communication). 
Column 3 gives the total amount of light produced by each class; 
Column 4 gives the fractional contribution of each class to the 
total luminic flux of the island.}
\begin{tabular}{cccc}
               &                     &         &              \\
\hline
Type of lamp   &  Number             & MLumens & Fraction     \\
               &                     &         & of total Flux \\
\hline
LPS          & $11086$             & $72000$ & $0.45$      \\
HPS           & $1350$              & $35000$ & $0.22$       \\
Mercury      & $1040$              & $14800$ & $0.09$      \\
Incandescent & $1026$              & $30150$ & $0.19$    \\
Fluorescent$~$compact  & $560$     & $670$   & $<0.01$   \\
Tube-discharge & $2104$            & $6312$  & $0.04$    \cr
\hline
\end{tabular}
\end{center}
\end{table*}

\section{Observational data}
Our sky spectra were obtained from archival science frames taken 
in the period August-December $2003$ with the $3.58$m Telescopio 
Nazionale Galileo at La Palma using DoLoRes (Device Optimized for 
Low Resolution), equipped with a $2048 \times 2048$ pixel thinned 
back-illuminated CCD with $15\mu$ pixels. 
Only spectra taken with the LR-B Grism were considered, 
with a final wavelength coverage of $\sim 3800 \div 8000 \AA$. 
The slit widths used were $1.0"$ and $1.3"$, yielding a 
resolution of $2.8 \AA /$pix and $3.6 \AA /$pix respectively. 
Wavelength comparison lines were obtained with a Helium lamp at the beginning 
of each night. 
For the present study, only deep exposures taken with airmass $< 
1.3$ during photometric, moonless nights with low extinction were selected. 
After a careful visual inspection, those spectra showing very similar 
content of light pollution lines were aligned and co-added to build 
six template spectra (hereinafter groups). These groups span a wide range in azimuth, 
epoch of the year and observing conditions, crucial to disentangle 
environmental and seasonal effects. 
As reported by BE98, we also found noticeable night-to-night variations in the 
intensity of the light pollution lines; this could be due to the presence of clouds 
below the ORM, blocking most of the outdoor lighting. To reduce the errors 
on the 
final line fluxes, we decided to include in the same 
group only those spectra whose NaD$_{5892}$ line fluxes differred by no 
more than $30\%$.
In particular, the spectra with the highest Na line fluxes (less cloud 
cover) were considered. 
Our data were reduced using standard IRAF tasks for long-slit spectra. 
The final wavelength calibration is accurate to $\sim 0.8~\AA$ r.m.s. 
Flux calibration was performed by observing spectrophotometric 
standard stars (typically one per night); within each group, the 
individual response functions were averaged to reduce errors introduced by 
slit losses and the variability of the photometric quality of the nights. 
The final flux calibration is accurate to $\sim 15\%$.

%
\section{Analysis of the night sky spectra}

\begin{table*}
\begin{center}
\caption[]{Overall properties of the night sky spectra. Last column 
reports when the exposures were taken (either before or after local midnight, when 
restrictions to the outdoor lighting of La Palma are applied; see Sect.2).}
\begin{tabular}{ccccccc}
       &            &         &                &         &        &      \\
\hline
Name   & Exposure & Slit Width & Azimuth & Airmass  & Extinction  & Date \& Time\\
       & Time (h)       & (arcsec)   & (degrees) &        & r-band (mag)  &    \\
\hline
Group1 & $4.0$      & $1.0$   & $+90 \div 250$ & $1.19$  & $0.1$  & 
$Jul$-$Aug2003;After~h24$  \\
Group2 & $2.0$      & $1.0$   & $+265$         & $1.25$  & $0.11$ & 
$29/08/03; Before~h24$ \\
Group3 & $1.0$      & $1.0$   & $+215$         & $1.30$  & $n.a.$ & 
$29/10/03; Before~h24$ \\ 
       &            &         &                &         &        & 
$+thin~clouds$\\
Group4 & $1.0$      & $1.3$   & $+154$         & $1.21$  & $0.12$ & 
$27/12/03; Before~h24$ \\
Group5 & $1.5$      & $1.3$   & $+170$         & $1.30$  & $0.12$ & 
$27/09/03; Before~h24$  \\
Group6 & $2.0$      & $1.3$   & $+181$         & $1.15$  & $0.14$ & 
$27/09/03; After~h24$  \\
\hline
\end{tabular}
\end{center}
\end{table*}

%
\begin{table*}
\begin{center}
\caption[]{Fluxes of the most important emission lines as measured in our 
spectra.Values are in Rayleigh (see BE98 for some useful conversion formulas). 
When not detected, a line is labeled with "n.d."; if the 
line was too noisy/faint or either blended with another line, it is labeled with 
"n.a.". Contribution to NaD$_{5890-6}$ from light pollution is shown in 
parentheses (see Sect 4.1 for details).}
\begin{tabular}{ccccccc}
       &            &         &                &         &        &      \\
\hline
Line     & Group$1$ & Group$2$ & Group$3$ & Group$4$  & Group$5$  & Group$6$  \\
\hline
Hg$_{4046}$    & $3.4$    & $5.2$   & $9.5$      & $6.1$   & $10.2$  & $6.3$  \\
Hg$_{4358}$    & $5.6$    & $7.9$   & $22.0$     & $17.6$  & $14.2$  & $4.5$ \\
NI$_{5199}$    & $1.5$    & $15.4$  & $3.2$      & $11.2$  & $5.1$   & $3.1$ \\
Hg$_{5461}$    & $4.4$    & $5.5$   & $25.7$     & $10.9$  & $8.6$   & $4.7$ \\
OI$_{5577}$    & $310$    & $256$   & $303$      & $234$   & $447$   & $504$  \\
Na$_{5683-8}$  & $3.5$    & $6.3$   & $30.6$     & $9.5$   & $11.4$  & $3.6$  \\
Hg$_{5769}$    & $n.d.$   & $n.d.$  & $7.2$      & $n.d.$  & $1.9$   & $1.4$  \\
Hg$_{5790}$    & $n.d.$   & $n.d.$  & $4.7$      & $n.d.$  & $1.7$   & $0.7$  \\
NaD$_{5890-6}$& $189(156)$ & $148(89)$& $658(431)$ & $284(134)$ & $251(162)$   
& $270(161)$  \\
Na$_{6154-61}$ & $n.a.$   & $n.d.$  & $9.5$      & $n.a.$  & $9.6$   & $n.a.$ \cr
\hline
\end{tabular}
\end{center}
\end{table*}

\subsection{Na Lines - natural and artificial contributions}

Given the population of lamps at La Palma, the NaI lines are by 
far the most important sources of light pollution at ORM. BE98 reported a median 
equivalent width of NaD$_{5892}$ of $\sim 100 \AA$ ($\sim 100$R) during 
summer, of which $\sim 70$R due to outdoor lighting and $\sim 30$R due to 
the 
natural skyglow. The natural NaD skyglow is known to have a strong seasonal variation, 
going 
from $\sim 30$R in summer to $\sim 200$R in winter (Schubert \& Walterscheid 2000).
A noticeable effect we found in our spectra is the decrease of the Na and Hg lines 
in the spectra taken after local midnight, when most of the HPS and 
mercury lamps are switched off, according to the Canary Sky Law (see Sect. 2). 
\noindent
To disentangle the natural and artificial contributions to the NaD$_{5892-6}$ 
emission 
we used our Group5 and $6$ spectra taken respectively before and 
after midnight. Note that no seasonal effect is present since both of them were 
taken at the end of September 2003. 
We assumed that all the Na$_{5683-8}$ flux of Group6 is due to LPS lamps 
while that of Group5 is the sum of LPS and HPS contributions. Thus the fractional 
contribution of LPS to Na$_{5683-8}$ emission of Group5 is $3.6/11.4 = 0.32$ and that 
of HPS 
is $0.68$. From the Philips catalog of lamps we derived the ratio 
NaD$_{5892-6}$ / Na$_{5683-8}$$= 44.6$ for the SOX LPS $35$W lamps 
mostly used at La Palma. For Group6 we calculate that light pollution 
from LPS lamps contributes $\sim 3.6 \times 44.6 = 161$R to the 
NaD$_{5890-6}$ flux; Group5 has an identical value since LPS lamps are never 
switched off during the night. We deduce that at the end of September $2003$ the 
natural NaD$_{5892-6}$ skyglow at ORM was $\sim 90-100$R.

\noindent
We also tried another approach to verify our assumptions about the fluxes of 
Na$_{5683-8}$ for Groups $5$ and $6$. 
From Table 2 the ratio of the illumination contribution of HPS vs. LPS 
lighting in La Palma is $\sim 0.48$. 
From the Philips catalog of lamps, as most of the HPS lamps at La 
Palma are SON-T $70$W, we calculate that the flux of Na$_{5683-8}$ 
emitted by a LPS lamp is $0.38$ times that emitted by a HPS lamp. 
Thus, for Na$_{5683-8}$ of Group5 we obtain that $3.4$R are from LPS lamps and $8.0$R 
are from HPS lamps.
These values are in very good agreement with those obtained above by simply 
assuming that all the flux of Na$_{5683-8}$ in Group6 ($3.6$R) comes from LPS lamps.

\noindent
Group1 (see Fig.1) is our longest exposure spectrum and well represents the 
average 
observing conditions at ORM after midnight when looking at $\pm 5$hrs from the 
meridian. 
The first important difference from BE98 is that we now clearly detect 
Na$_{5683-8}$ emission, while Na$_{6154-61}$ is still undetected. 
Moreover, the Group1 spectrum shows that the average contribution of light 
pollution to the NaD$_{5892-8}$ flux in the southern regions of sky 
after midnight is $\sim 150$R, about twice the value measured in $1998$.

\noindent
Group2 (see Fig.1) is interesting because it was taken towards the NW, a 
zone with relatively low light pollution (see Table 1) as confirmed by the lowest 
contribution 
of artificial NaD$_{5892-8}$ detected in our spectra ($89$R). With respect to Group1, 
the higher flux of Na$_{5683-8}$ is due to the fact that Group2 was taken 
before local midnight. 

\noindent
Group3 has light pollution lines with abnormally high fluxes (see Fig.1). 
It was taken looking in the direction of 
the most polluting towns of the island, before midnight and with thin clouds above 
the ORM (no data are available for the atmospheric extinction). 
A direct estimate with the above explained procedure of the artificial contribution 
to the NaD$_{5892-8}$ gives 
$431$R, which would result in a natural NaD background of $227$R, somewhat 
higher than expected at the end of October. 
In this case, the presence of high clouds could have played a role in reflecting 
back light pollution to the observatory. 

\noindent
Group4 is a typical spectrum taken looking toward a moderately polluted 
region of sky, $\sim 2$hrs before meridian. Here, the effects of the two urban 
areas of Brena Alta/Brena Baja and partly of Santa Cruz de La Palma (see Table 
1) are evident. The higher-than-average levels of the Na lines 
(note the Na$_{5683-8}$ flux of $9.5$R) are 
also due to the fact that it was taken before midnight. We estimate 
the contribution of light pollution to the NaD$_{5892-8}$ to be $134$R. 

\noindent
The above discussed Group5 and Group6 are typical spectra taken at the 
meridian where the line of sight intercepts the town of El Paso (see Table 1). 
The decrease of the Na lines fluxes is evident in Group6, taken after 
midnight. The contribution of light pollution to the NaD$_{5892-8}$ is $\sim 160$R, 
similar to that of Group4 and Group1.

From Table 4 it is evident that in all our spectra, the 
fluxes of the NaD$_{5892-6}$ line are always $1.5 - 2.5$ times higher than those 
of BE98. 
In principle this indicates that light pollution due to LPS and HPS lamps 
considerably increased in the last $5$ years at La Palma, despite the efforts 
made to control it.

\subsection{Hg I Lines}

If we consider Group1, the emission of the 
lines Hg$_{4358}$ and Hg$_{5461}$ is about half that reported in 
BE98 but our 
spectrum also shows the line Hg$_{4046}$ detected for the first time at ORM and with 
intensity comparable to Hg$_{5461}$. 

\noindent
Although the Group2 spectrum was taken in a less polluted region of sky, it has 
$\sim 40\%$ more Hg emission than Group1 and half the Hg 
emission of Groups $4$ and $5$ taken toward two towns before midnight 
(see Tables 3 and 4). This demonstrates the benefits of the Canary Sky Law; 
observations made in the less polluted region of sky before midnight imply 
higher fluxes of Hg lines than those made toward a more polluted region but after midnight.

\noindent
The most striking feature in our spectra is the line detected in Group3 
(see Fig.1) 
at $5355.5 \AA$ which we identified as ScI (tabulated lambda is 
$5356.09 \AA$, see Table 6 of 
Slanger et al. 2003). Scandium is used as an additive to high-pressure metal 
halide lamps. Since on La Palma these are used only in the soccer 
stadiums (to be estinguished after 23:00), our detetection could 
have coincided with some nocturnal sporting activity. 
The line at $5351.1 \AA$ detected in Group4 (see Fig.2) can also be 
identified as ScI 
emission (tabulated lambda $5349.71 \AA$). The Group3 shows other two lines 
never detected before 
at ORM: Hg$_{5769}$ and Hg$_{5790}$, only observed at Mount Hamilton (Slanger 
et al. 2003) and Kitt Peak (Massey et al. 1990). 
Though very faint, these lines also appear in our Groups $5$ and $6$, with a 
clear dimming after midnight evident in the latter spectrum (see Table 4).

\noindent
To conclude, the average fluxes of the Hg lines detected in our spectra are $\sim 
50\%$ fainter than those reported in BE98. When observing toward a town, the 
Hg lines have about the same intensities as in 1998.
Our directional spectra show for the first the effect of the application of the 
Sky Law after midnight but it is evident that mercury lamps are 
never completely extinguished after that hour, since Hg lines are present in all 
our spectra. 
For a typical town like El Paso (see Group5-6), we infer that only half of the 
mercury lamps are estinguished after midnight.

\section{Spectrophotometry}
Synthetic night sky brightness measures at 
B, V and R bands were also calculated from  our spectra. 
The advantage of using spectra is that both natural airglow (OI$_{5577}$) 
and artificial (NaD$_{5892-8}$) 
lines can be eliminated by replacing them with the average continuum. 
It is noteworthy that the OI$_{5577}$ line typically contributes 
$0.16$ mag arcsec$^{-2}$ to the broadband V (Massey and Foltz 2000). 
The values presented in Table 5 were obtained as in Massey and Foltz 
(2000) and reported to zenith as in BE98. 
The contribution (in mag arcsec$^{-2}$) of the NaD$_{5892-8}$ 
emitted by LPS lamps to the V and R magnitudes is also reported in Table 5.

\noindent
The sky brightness values in Table 5 are consistent with those 
of BE98 if we take into account that our observations were made about seven  
years after the $1996.5$ solar minimum (sky is $\sim 0.4$mag 
darker at solar minimum). 
From Table 5 it is evident a significant increase of light pollution 
from the NaD$_{5892-8}$ line emitted by LPS lamps in La Palma (BE98 
reported a $0.05$mag contribution).

\begin{table*}
\begin{center}
\caption[]{Synthetic sky brightness measures (mag arcsec$^{-2}$) as 
obtained from 
our spectra (see Section 5). The natural OI$_{5577}$ and artificial 
NaD$_{5892-8}$ emission were replaced by the average continuum. 
LPS-V and LPS-R 
indicate the contribution of NaD$_{5892-8}$ (in mag arcsec$^{-2}$) emitted 
by low-pressure sodium lamps to the V and R magnitudes respectively.}
\begin{tabular}{cccccc}
           &           &        &       &       &   \\
\hline
Spectrum  & B	    & V	      &  R      & LPS-V  & LPS-R  \\
\hline
Group1	  & $22.48$ & $21.66$ & $20.74$ & $0.09$ & $0.11$ \\
Group2	  & $22.46$ & $21.74$ & $20.79$ & $0.05$ & $0.07$ \\
Group3	  & $22.34$ & $21.48$ & $20.47$ & $0.26$ & $0.31$ \\
Group4    & $22.40$ & $21.58$ & $20.69$ & $0.08$ & $0.10$  \\
Group5    & $22.40$ & $21.64$ & $20.72$ & $0.10$ & $0.12$  \\
Group6    & $22.42$ & $21.67$ & $20.77$ & $0.10$ & $0.12$  \cr
\hline
\end{tabular}
\end{center}
\end{table*}

\section{Conclusions}
Light pollution lines at the 
Roque de los Muchachos Observatory (ORM) - La Palma were studied with 
archive low-resolution spectra taken with DoLoRes at the 3.58m Telescopio 
Nazionale Galileo during 2003. Our spectra cover the 
wavelength range $\sim 3800 \div 8000 \AA$, 
and have resolution of $2.8 \AA /$pix and $3.6 \AA /$pix 
(slit width $1.0"$ and $1.3"$ respectively). Only deep exposures taken with 
airmass $< 1.3$ during photometric, moonless nights with low extinction were 
selected, resulting in six deep spectra which span a wide range in azimuth, 
epoch and observing conditions. 
We showed in Sect.4.1 how the detection of Na$_{5683-8}$ permits 
the artificial and the natural contributions to the NaD$_{5892-8}$ 
line to be disentangled. 
The average intensity of the NaD$_{5892-8}$ line emitted by LPS lamps 
increased by a factor of $1.5 - 2$ 
over the last $5$ years on La Palma and its contribution to the sky background 
is $0.05 - 0.10$mag at V-band and $0.07 - 0.12$mag at R-band, 
depending on the 
region of sky and the time when observations are made. The IAU's 
recommendation that NaD$_{5892-8}$ emission should not exceed in intensity 
the natural background, is definitely no longer met in La Palma. 
Sodium lines such as 
Na$_{5683-8}$ and Na$_{6154-61}$ were also detected in our spectra for the 
first time. 
Light pollution from mercury lamps is $\sim 50\%$ lower 
than in 1998, except when observations are made looking toward the towns, before 
midnight; in this case we found very similar levels. 
Our spectra also show the Hg$_{4046}$ and, in two cases, the Hg$_{5769}$ and 
Hg$_{5790}$ lines, never detected before at ORM.
Though in non-optimal atmospheric conditions, we detected in Group3 one 
strong line which was identified as Scandium (ScI). This element is used as 
an additive in 
high-pressure metal halide lamps which, to our knowledge, are only used in the 
soccer stadiums on La Palma. 
The presence of this type of lamp on La Palma is confirmed by 
another line at $5351.1 \AA$ detected in the Group4 spectrum which can also be 
identified as ScI emission. 
\noindent
Synthetic sky brightness measures were derived from our spectra at V, B 
and R bands (see Section 5). Our values are in good agreement with those 
of BE98 if we take into account that our observations were done at 2003, 
about seven years after the last sunspot minimum (sky is $\sim 0.4$mag
darker at solar minimum). 

\section{Acknowledgments}
The author is particularly grateful to Dr. Javier Francisco Diaz Castro, of the 
O.T.P.C. (Oficina tecnica para la Proteccion del Cielo-IAC) for 
providing updated data of the outdoor lighting of La Palma and for useful 
hints 
for the manuscript. The author is also grateful to Dr. Chris Benn (Isaac Newton 
Group-La Palma) and Dr. William Cochran (McDonald Observatory) for their 
helpful suggestions and discussions. 
Based on observations made with the Italian Telescopio Nazionale Galileo (TNG) 
operated on the island of La Palma by the Centro Galileo Galilei of the INAF 
(Istituto Nazionale di Astrofisica) at the Spanish Observatorio del Roque de los 
Muchachos of the Instituto de Astrofisica de Canarias.

\begin{figure}[here]
\includegraphics*[scale=0.6,angle=-90.0]{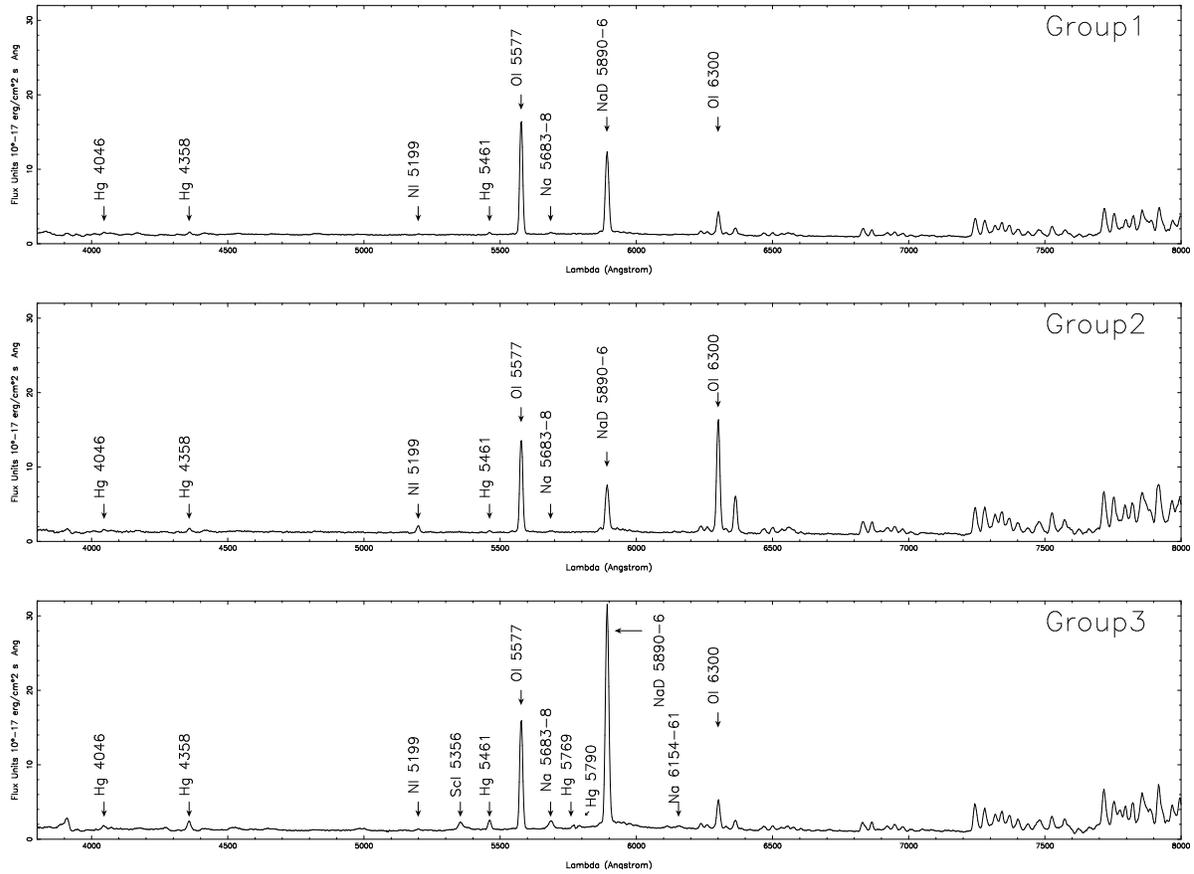}
\caption{The night-sky spectra (see Table 3). The Group1 ($4$hrs total 
exposure) is the average of $8$ spectra and best represents the average 
observing conditions at ORM; The Group2  
spectrum was taken towards the NW, the least light-polluted zone at ORM. 
The Group3 spectrum was taken towards the most light-polluted 
region of sky at ORM, before midnight. The presence of thin clouds could 
explain the abnormally high fluxes of the light polluting lines (see 
Sect.4.1).}
\end{figure}
%
%
\begin{figure}[here]
\includegraphics*[scale=0.6,angle=-90.0]{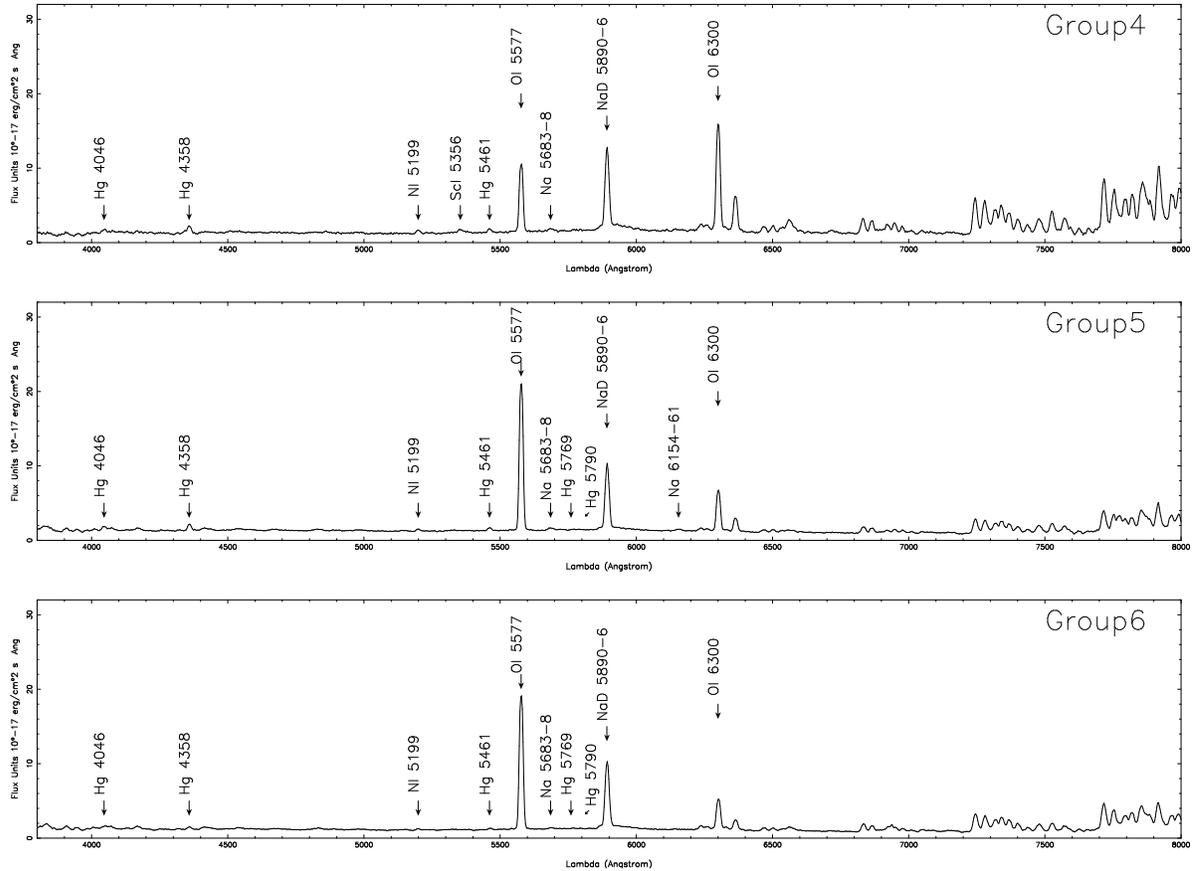}
\caption{The night-sky spectra (see Table 3).
The Group4 spectrum was taken toward a moderately polluted 
region (see Table 1) before midnight; The Group5 
spectrum was taken toward the meridian before midnight;
The Group6 spectrum was taken toward the meridian after midnight.}
\end{figure}


\begin{thebibliography}{}

\bibitem{BE98}
Benn, C.R., Ellison, S.L. 1998, La Palma Technical Note, 115

\bibitem{Ga89}
Garstang, R.H. 1989, PASP, 101, 306 
\bibcode{1989PASP..101..306G}

\bibitem{Ma90}
Massey, P., Gronwall, C., Pilachowsky, C.A. 1990, PASP 102, 1046
\bibcode{1990PASP..102.1046M}

\bibitem{Ma00}
Massey, P., Foltz, C.B., 2000, PASP 112, 566
\bibcode{2000PASP..112..566M}

\bibitem{MN94}
McNally, D. 1994 ed., "The Vanishing Universe - Adverse Environmental Impacts 
on Astronomy", Cambridge University Press.

\bibitem{M73}
Munoz-Tunon, C., Vernin, J., Varela, A.M.  1997, A\&AS, 125, 183
\bibcode{1997A\&AS..125..183M}

\bibitem{SW00}
Schubert, G., Waltersheid, R.L., 2000, in Allen's Astrophysical Quantities, 
ed. A.N. Cox (New York: AIP Press; Springer), 4$th$ edition.

\bibitem{Sl03}
Slanger, T.G., Cosby, P.C., Osterbrock, D.E. et al. 
2003, PASP, 115, 869
\bibcode{2003PASP..115..869S}
\end{thebibliography}
\end{document}